\pdfoutput=1
\documentclass[aps,prl,preprint,showpacs,superscriptaddress]{revtex4}
\usepackage{amsmath}
\usepackage{dcolumn}
\usepackage{epsfig}
\usepackage{graphicx}
\usepackage{latexsym}
\usepackage{amssymb}
\usepackage{times}
\usepackage{natbib}    
\begin{document}

\title{Tetragonal CuO: A new end member of the 3d transition metal monoxides}

\author{Wolter Siemons}
\affiliation{Faculty of Science and Technology and MESA+ Institute for Nanotechnology, University of Twente, 7500 AE
Enschede, The Netherlands} \affiliation{Geballe Laboratory for Advanced Materials, Stanford University, Stanford,
California 94305, USA}
\author{Gertjan Koster}
\affiliation{Faculty of Science and Technology and MESA+ Institute for Nanotechnology, University of Twente, 7500 AE
Enschede, The Netherlands} \affiliation{Geballe Laboratory for Advanced Materials, Stanford University, Stanford,
California 94305, USA} \email{g.koster@utwente.nl}
\author{Dave H. A. Blank}
\affiliation{Faculty of Science and Technology and MESA+ Institute for Nanotechnology, University of Twente, 7500 AE
Enschede, The Netherlands}
\author{Robert H. Hammond}
\affiliation{Geballe Laboratory for Advanced Materials, Stanford University, Stanford, California 94305, USA}
\author{Theodore H. Geballe}
\affiliation{Geballe Laboratory for Advanced Materials, Stanford University, Stanford, California 94305, USA}
\author{Malcolm R. Beasley}
\affiliation{Geballe Laboratory for Advanced Materials, Stanford University, Stanford, California 94305, USA}

\begin{abstract}
Monoclinic CuO is anomalous both structurally as well as
electronically in the 3$d$ transition metal oxide series.  All the
others have the cubic rock salt structure.  Here we report the
synthesis and electronic property determination of a tetragonal
(elongated rock salt) form of CuO created using an epitaxial thin
film deposition approach. In situ photoelectron spectroscopy
suggests an enhanced charge transfer gap $\Delta$ with the overall
bonding more ionic. As an end member of the 3d transition
monoxides, its magnetic properties should be that of a high $T_N$
antiferromagnet.
\end{abstract}


\maketitle

\section{Introduction}
\begin{figure}[tb]
\centering
\includegraphics[width=1\columnwidth]{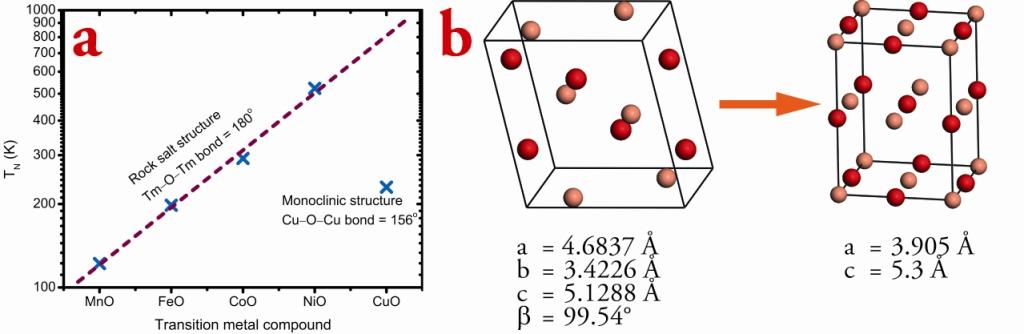}
\caption{(a) An overview of the N\'{e}el temperatures for
transition metal monoxides with a rock salt structure. $T_N$
increases exponentially with each element up to NiO as indicated
by the dashed line, which is a fit of the first four points. The
$T_N$ of the monoclinic CuO structure does not follow this trend,
but is much lower, possibly due to a change in the Tm--O--Tm bond
angle. When a rocksalt structure of CuO is formed $T_N$ might
follow the trend and linear extrapolation predicts a $T_N$
slightly higher than 800 K. (b) The change in unit cell symmetry
going from bulk (monoclinic) to the strained unit cell on
SrTiO$_3$ (tetragonal), including the unit cell parameters. The
lighter colored atoms represent oxygen an the darker colored atoms
the copper.} \label{figure1}
\end{figure}

Since the discovery of high temperature superconductivity
in the copper oxide perovskites, its origin and mechanism are
still unexplained and under debate. The original proposal by
Bednorz and Mueller \cite{bednorz1988} and others, that a
Jahn-Teller distortion in a highly symmetric divalent copper
monoxide structure introducing a strong electron-phonon
interaction causes superconductivity, has lead to extensive
studies toward the synthesis of the family of cuprates. The
simplest form, the cubic rock salt copper monoxide, however, has
not been found in nature nor has been successfully synthesized,
yet. CuO is the exceptional member of the rock salt series as one
traverses the periodic table from MnO to CuO. It deviates
substantially from the trends exhibited by the members with lower
atomic number. All the others have the cubic rock salt structure
and all are correlated antiferromagnetic insulators.
\cite{mattheiss1972,terakura1984,harrison2007,zaanen1987} CuO
differs in having a monoclinic structure as opposed to the rock
salt structure of the other monoxides, and, as shown in Fig.
\ref{figure1}, it also has a substantially lower N\'{e}el
temperature than a simple extrapolation of the trend across the
periodic table would suggest. Presumably, this exceptional
behavior is a consequence of its lower symmetry structure.
Clearly, the properties of CuO in higher symmetry structures would
be of great fundamental interest in understanding correlated
materials.\cite{georges1996,ahn2003} Here we report the synthesis
and preliminary electronic property determination of a tetragonal
(elongated rock salt, displayed in Fig. \ref{figure1}b) form of CuO
created for the first time, by using an epitaxial thin film
deposition approach. The results demonstrate that higher symmetry
phases of this important correlated oxide are possible and now
available for physical studies. Looking ahead, if the trend shown
in Fig 1a were followed, the N\'{e}el  temperature of rock salt
CuO would be very high (700 to 800 K), as would be its associated
exchange coupling $J$. If such a high-$J$ CuO could be doped, its
properties would be of great interest in the context of the
earlier mentioned high-T$_{\mathrm{c}}$ superconductors.

 The films discussed in this work have
been well characterized from the structural point of view using reflection high energy electron diffraction (RHEED) and
x-ray photoemission diffraction (XPD), and from the electronic structure point of view using in-situ x-ray and
ultraviolet photoemission spectroscopy ((XPS/UPS). As we will discuss below, by comparing with the stable copper
oxides, we have determined the Cu has a charge of +2 and that the tetragonal structure is more ionic than the
monoclinic phase. Preliminary attempts to dope our films using charge transfer from over-layers have also been carried
out.   Taken together, the results demonstrate that higher symmetry forms of this important correlated oxide are
possible and available for physical study.

\section{Synthesis and Analysis}

Our epitaxial films of tetragonal CuO were grown in a UHV chamber
on SrTiO$_3$ substrates using pulsed laser deposition (PLD), although we have also grown
sample using electron beam evaporation. For PLD a Lambda Physic LPX 210 KrF excimer laser produces a 248 nm wavelength
beam with a typical pulse length of 20--30 ns. A rectangular mask shapes the beam selecting only the homogeneous part,
and a variable attenuator permits variation of the pulse energy. The variable attenuator also offers the possibility to
run the laser at the same voltage every run, which ensures the same pulse shape every time and therefore the best
reproducibility. A lens makes an image of the mask on the target resulting in a well defined illuminated area. Energy
density on the target was kept at 2.1 J cm$^{-2}$ with a repetition rate of 1 Hz. The temperature of the substrate was
fixed to 600 $^{\circ}$C. Atomic oxygen was provided during deposition by a microwave plasma source (Astex SXRHA) with
the source operating at 600 W with a flow of 2.5 sccm oxygen, resulting in a background deposition pressure in the
system of 1.5 $\times$ 10$^{-5}$ Torr and an estimated 3 $\times$ 10$^{17}$ oxygen atoms cm$^{-2}$
s$^{-1}$.\cite{ingle1999} The growth was monitored by RHEED.

Films were grown on either insulting or conducting
(0.5\% Nb doped)  SrTiO$_3$. The latter offers the advantage of
reduced charging when performing photoemission spectroscopy (PES)
on the insulating CuO samples. Both types of substrates were
TiO$_2$ terminated, as described by Koster and
coworkers.\cite{koster1998}  In order to stabilize the new phase,
it is of utmost importance to oxidize Cu to a 2+ state. This was
accomplished by using a target of CuO and providing atomic oxygen
during deposition.

After deposition samples were cooled down in atomic oxygen. The films were typically unstable in atomic oxygen below
300 $^{\circ}$C.  Specifically, it was found that the films would relax to tenorite when cooled to room temperature
under deposition conditions. To avoid this problem,  the atomic oxygen was switched off at 300 $^{\circ}$C and the
sample then cooled to room temperature in molecular oxygen ($\sim$10$^{-5}$ Torr).  These results imply that the
stability line for the epitaxially stabilized tetragonal CuO when exposed to atomic oxygen lies around 300
$^{\circ}$C\@.

The thickness of the tetragonal CuO samples is limited by a relaxation to the tenorite phase above a certain thickness.
For most samples 300 laser pulses were used in the PLD process to guarantee a streaky RHEED pattern with no 3D spots.
This corresponds to a layer thickness between 15 and 20 {\AA} as determined with x-ray reflectivity and angle resolved
XPS measurements.\cite{spruytte2001} AFM imaging confirms that thin samples are flat with the SrTiO$_3$ step structure
still visible, whereas the samples that are thicker exhibit islands on top, associated with the film growth process,
that cause the observed 3D RHEED pattern.

Let us note here, that after exposure to air the top layer of the CuO was found to degrade to tenorite, whereas the
layer closest to the interface with SrTiO$_3$ is found to be mostly tetragonal. This result could be established using
angle dependent XPS, as discussed in detail in the thesis by Siemons.\cite{siemonsthes2008}   Also, when a sample is
kept under vacuum after growth, the CuO is slowly reduced over a period of days.\cite{shen1990} After four days the
intensity of the satellite peaks are reduced to about half their size after deposition. All XPS spectra are referenced
with respect to the Ti 2$p$ peak of the substrate.

The XPS and UPS measurements are performed \emph{in situ} with a VG scientific ESCAlab Mark II system. The photon
source for XPS is Al k$\alpha$ and for UPS measurements HeI (21.2 eV) radiation was used. Both sources are
non-monochromatic and spectra are corrected for satellites by use of software.

\section{Initial Growth}
When grown on doubly terminated substrates, the growth was found
to be more 3D, and the films relaxed to the tenorite phase at an
earlier stage in the deposition. Similar results were observed on
other substrate materials, such as DyScO$_{3}$ and LaAlO$_{3}$,
where 3D growth patterns of the relaxed structure would occur at a
very early stage. Due to the polar nature of these materials those
substrates are either doubly terminated (a reliable method to make
them singly terminated such as with SrTiO$_3$ has not been fully
developed), or complicates the layer-by-layer growth of a neutral
material such as CuO \cite{nakagawa2006}. In addition, these other
substrates have different lattice parameters as well, which might
affect their efficacy for epitaxy of CuO.

\begin{figure}[tb]
\centering
\includegraphics[width=1\columnwidth]{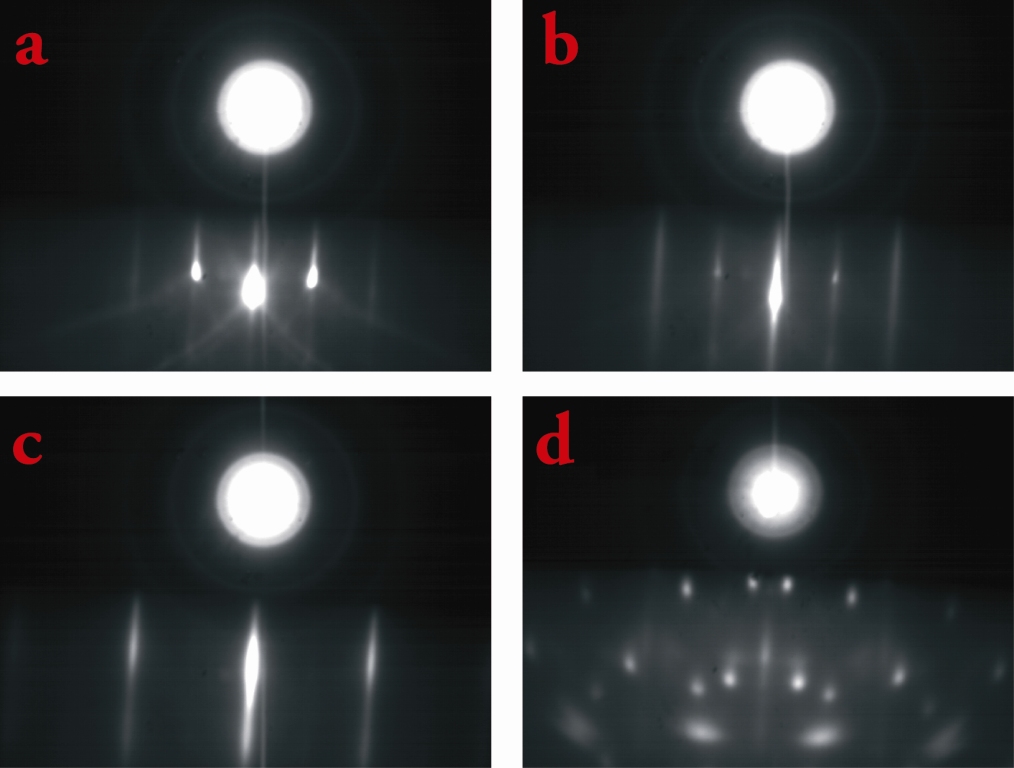}
\caption{(a-c) The evolution of the RHEED spectrum during the
growth of CuO on SrTiO$_3$ taken along the (01) direction of
SrTiO$_3$ shows a transition from a clear 2D pattern for bare
SrTiO$_{3}$ (a) to the streaky pattern of CuO (c). (d) When growth
is continued the film relaxes to tenorite showing clear 3D spots.}
\label{figure2}
\end{figure}

Fig. \ref{figure2}a--c show the evolution of the RHEED pattern in
the case of successful growth of the tetragonal phase.
Specifically, a streaky pattern, which is 4-fold symmetric when
the sample is rotated around the surface normal, emerges during
deposition, without any 3D spots between the streaks.  In
contrast, Fig. 2d shows the RHEED pattern that emerges when the
film relaxes to the tenorite phase.  It is clearly different from
that of the new tetragonal phase. Such a streaky pattern was
observed before for the growth of CuO on MgO,\cite{catana1992} but
no different symmetry of the CuO unit cell was observed in that
work.

\section{Determining Lattice Parameters}
\begin{figure}[tb]
\centering
\includegraphics[width=1\columnwidth]{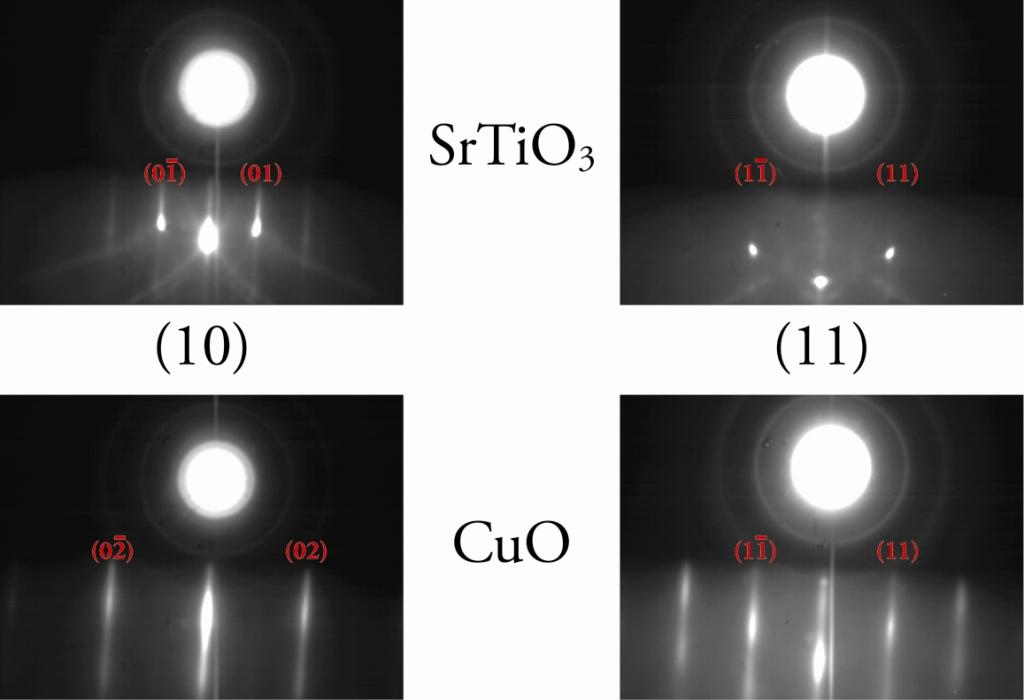}
\caption{The electron diffraction patterns of SrTiO$_3$ and CuO
taken along their (10) and (11) directions. Even though the (10)
reflection of CuO is not allowed as explained in the text, very
faint lines at its position are usually visible. The patterns
shown here are both fourfold symmetric when the sample is rotated
around its azimuthal axis.} \label{figure3}
\end{figure}

To determine the in-plane lattice parameters of the new phase, RHEED spectra were used. Spectra taken along the
SrTiO$_3$ (10) and (11) are shown in figure \ref{figure3}.  Since the lattice parameters of the SrTiO$_3$ are well
known (3.905 {\AA} cubic), the lattice parameters of CuO can be calculated by comparing its lines to those of
SrTiO$_3$.  There is a subtlety here, however. One must correctly identify the lines.   For a rock salt structure, not
all the diffraction peaks are allowed, and those that are allowed are different in 2D and 3D.  We find that for our
very thin films the 2D result is required.  Specifically, in 2D the structure factor becomes:

\begin{equation*}
\begin{aligned}
   \text{h,k unmixed: }  &F_{hk}=2\sum_{n/2} f_n\mathrm{e}^{2\mathrm{\pi}\mathrm{i}(hx_n+ky_n)}\\
   \text{h,k mixed: }    &F_{hk}=0\\
\end{aligned}
\end{equation*}

Therefore, in the 2D case, the (01) reflection is not allowed but the (11) is, and it follows that the shortest spacing
between the RHEED lines corresponds to the (11) reflection.  On the other hand, if the pattern were 3D, the shortest
spacing would correspond to the (01) reflection.  Analysis of the data assuming such an identification leads to
non-physical results.

With the dimensionality established, it is straightforward to determine the lattice constants.   In the (11) direction
of the SrTiO$_3$ substrate, the (11) and (22) reflections of CuO are allowed, and along the (01) direction of the
SrTiO$_3$, the (02) and (04) reflections of CuO are allowed. The CuO in-plane lattice parameters are exactly the same
size as the SrTiO$_3$ underneath. This corresponds to a Cu--O bond length of 1.95 {\AA} in-plane, a number which
corresponds well with values found in high T$_{\mathrm{c}}$ materials or in monoclinic CuO. If the unit cell were cubic
this would result in a unit cell volume of 59.5 {\AA}, too far from  the 81.1 {\AA}$^3$ of tenorite to be physically
reasonable. We conclude, therefore, that the unit cell must be elongated along the out-of-plane direction.  But based
on the RHEED spectra alone, we cannot draw conclusions about the out-of-plane lattice constant.

\begin{figure}[tb]
\centering
\includegraphics[width=0.7\columnwidth]{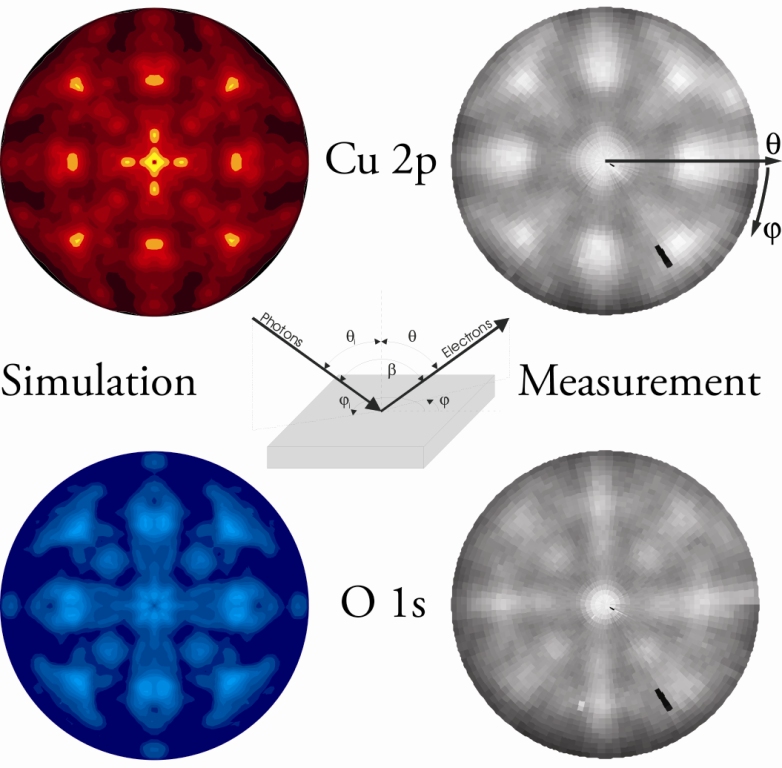}
\caption{XPD patterns for CuO measured (on the right) at two
binding energies: the O 1s and the Cu 2p$_{\frac{3}{2}}$ main
line. The measurements were performed with the pass energy set at
100 eV, an acceptance angle of 5 degrees in steps of 2$^{\circ}$
for $\theta$ (90-30$^{\circ}$) and 3$^{\circ}$ for $\phi$
(0-360$^{\circ}$). Simulations of the CuO in tetragonal form are
shown on the left, as calculated with software by Abajo \emph{et
al}.\cite{abajo2001} The Cu signal comes solely from the film,
whereas the O signal has a contribution from the substrate. The Cu
pattern shows clear four-fold symmetry, which is not what would be
expected from a single domain tenorite film.} \label{figure4}
\end{figure}

To measure the out-of-plane lattice constant, we employed XPD. To verify that the technique would produce
reliable results, it was first applied to a conducting Nb-doped SrTiO$_{3}$ substrate. Simulations and measurements
were strikingly similar. Two energies were measured: the O 1\emph{s} and the Cu 2\emph{p}. For the oxygen signal, a
contribution of the substrate is to be expected, since the escape depth of the electrons is of the order of the
thickness of the CuO layer. The copper signal, on the other hand, comes solely from the film because the substrate
contains no copper. The copper pattern shows a fourfold symmetry as shown on the right in figure \ref{figure4}, which
is indicative of a fourfold symmetric unit cell, such as a rock salt structure. By comparing simulations and
measurements the c-axis length of the unit cell was determined to be 5.3 {\AA}. Using the XPD data, the out-of-plane
lattice parameter cannot be determined more accurately. These lattice parameters are also shown in Fig. \ref{figure1}b.
Note that the films described in this work typically are 3 -- 4 unit cells thick.

With this complete set of unit cell dimensions, the unit cell volume now becomes 81.1 {\AA}$^3$, which is the same as
for tenorite and physically reasonable. The shortest Cu--O bond length in this structure is 1.95 {\AA}, which
corresponds to the value found in tenorite. The shortest Cu--Cu and O--O bond lengths are 2.76 {\AA}, a value that is
in between the values found for the two in tenorite.

\section{Electronic Structure}

\begin{figure}[tb]
\centering
\includegraphics[width=0.5\columnwidth]{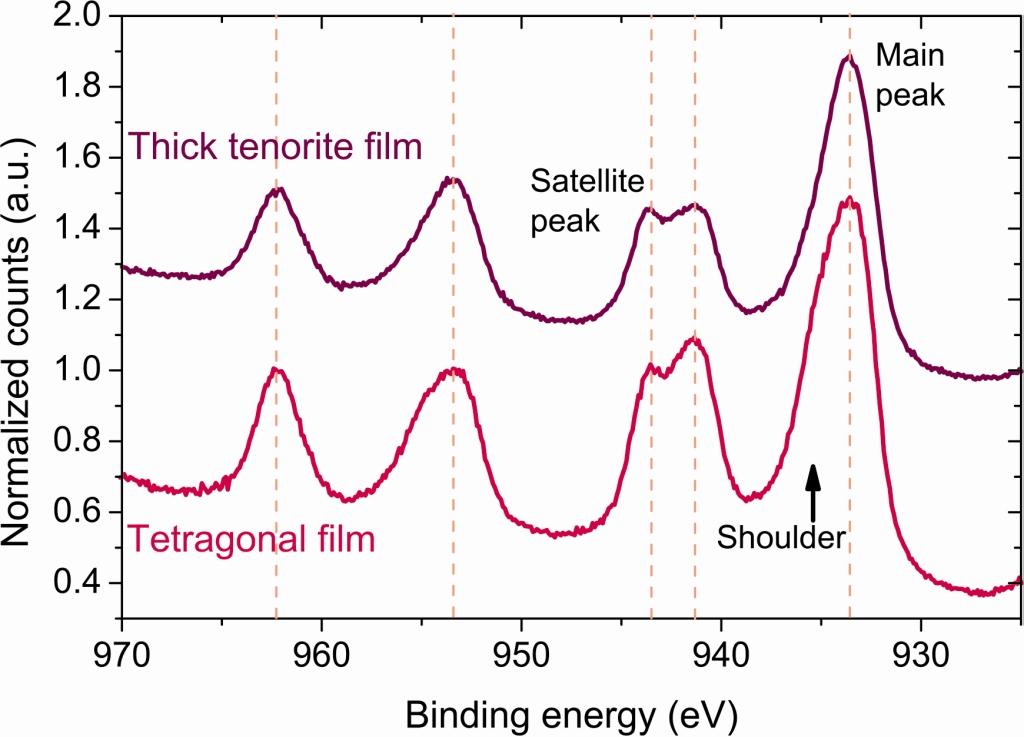}
\caption{The Cu 2\emph{p} core level XPS spectrum of the
tetragonal CuO looks significantly different from the tenorite
spectrum. The most striking differences are the broadening of the
main peak, the narrowing of the satellite peak, and the
redistribution of weight in the satellite peak. The spectra have
been normalized on the 2p$_{\frac{1}{2}}$ peak and the tenorite
spectrum has been given an offset for clarity.} \label{figure5}
\end{figure}

Let us now turn to the electronic structure of this new phase of CuO. The XPS spectra of the Cu 2\emph{p} lines for the
thin tetragonal epitaxial films as well as those for thicker films in the tenorite sturucture are shown in Fig.
\ref{figure5}.  For tenorite, we find 933.5 eV for the binding energy of the Cu 2\emph{p}$_{\frac{3}{2}}$ and 529.4 eV
for the O 1\emph{s}.  These correspond well to the values found in literature.  For our new tetragonal phase, the
energies of these lines do not appear to change significantly.  On the other hand, compared to the tenorite spectrum,
the tetragonal CuO spectrum has a larger shoulder on the main peak at higher binding energy, a narrower satellite peak,
and the spectral weight of the first satellite peak has shifted to lower binding energy. More precisely, when going
from the monoclinic to the tetragonal phase the full width at half maximum (fwhm) of the main peak increases from 3.6
to 4.1 eV, whereas the fwhm of the satellite peak decreases from 4.5 to 4.2 eV\@.

\begin{figure}[t]
\centering
\includegraphics[width=0.5\columnwidth]{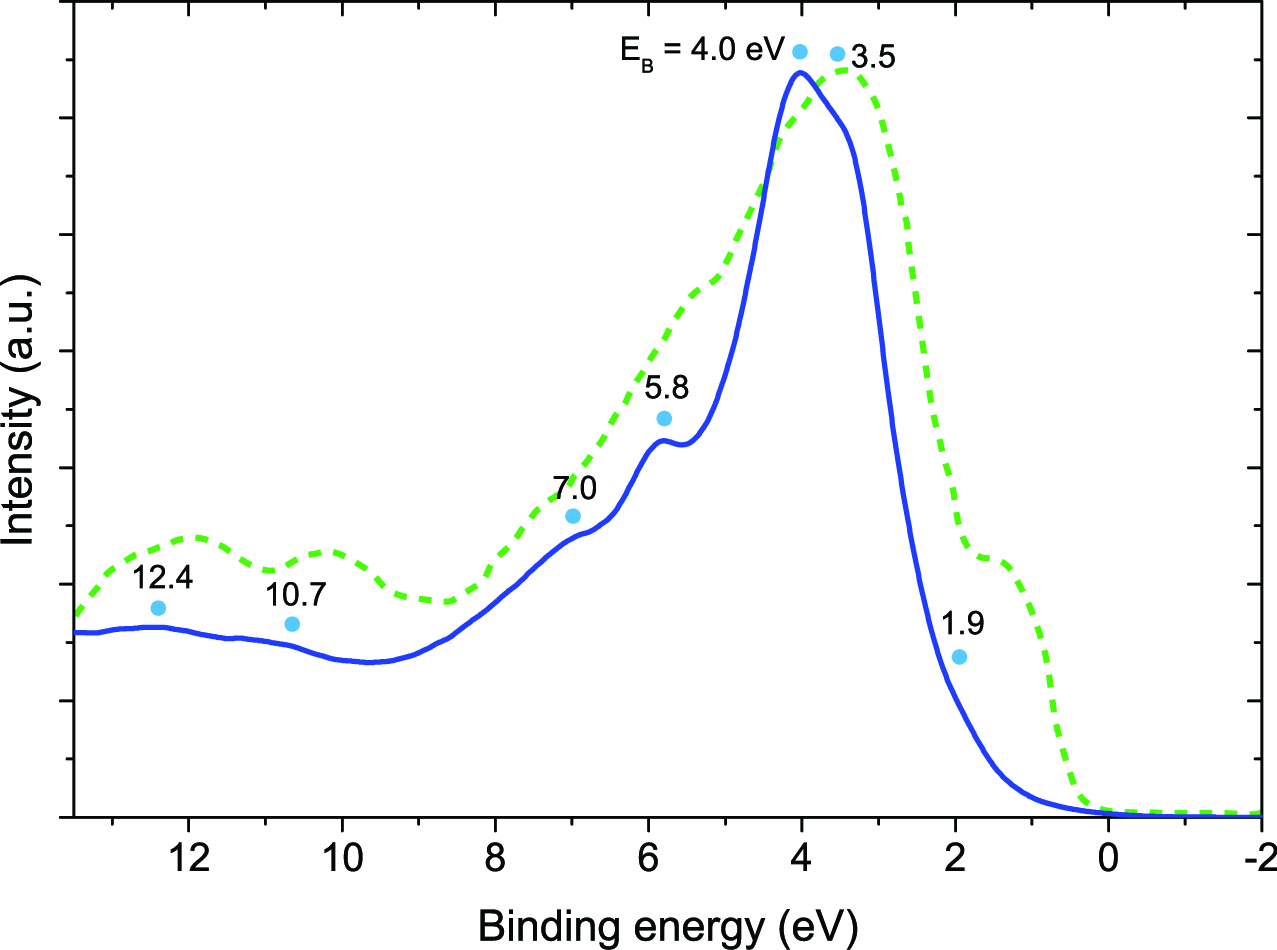}
\caption{The tetragonal CuO valence band spectrum, solid blue
line, measured with HeI radiation and the peak positions indicated
with cyan dots. The Nb doped SrTiO$_3$ substrate contributes to
the spectrum for such thin films, but its contribution is expected
to be very small. The dotted line is a typical tenorite spectrum
taken at 70 eV radiation, by Warren et al.\cite{warren1999}}
\label{figure6}
\end{figure}

The valence band spectrum was measured with He I radiation, and the results are presented in figure \ref{figure6},
together with a representative spectrum of tenorite, taken by Warren et al.\cite{warren1999}.  For the tenorite phase,
the spectrum is representative to that obtained by other
groups.\cite{ghijsen1988,shen1990,ching1989,ghijsen1990,warren1999} For example, there are no states at the Fermi
level, which is as expected for an insulator.

Comparing the spectra for tenorite and our CuO, there are some notable differences. The peak at low binding energy
($\sim$1.9 eV), which is caused by both oxygen and copper orbitals,\cite{takahashi1997} is not as strong in our
spectrum as it is in the spectra of tenorite. Also the largest peak, mainly belonging to copper, clearly consists of
two separate peaks in our spectrum, and is not observed before for tenorite. Note that calculations on the tenorite
band structure show a splitting of this peak\cite{ghijsen1988,takahashi1997}, and in some measurements a shoulder is
visible.\cite{ching1989} The peaks at 5.8 and 7.0 eV are both oxygen peaks and are also predicted by
calculations.\cite{takahashi1997} The double peak feature at higher energy (10--13 eV) is a copper feature only
predicted by the most advanced calculations.\cite{ghijsen1990,eskes1990,takahashi1997} All calculations on tenorite
predict the peaks to be at lower binding energies than measured here, and the spectrum we observe is shifted up by as
much as 2 eV. The spectrum for the tetragonal phase is shifted to higher binding energies by about 0.5 eV\@ compared to
the tenorite phase. Speculatively, assuming that the Fermi level is in the middle of the gap, this would imply a larger
charge transfer gap $\Delta$. For a film of this thickness, some contribution of the substrate to the spectrum is
expected. SrTiO$_3$ has two large peaks at higher energies (4.4 and 6.6 eV), but no clear evidence of a contribution of
these peaks and the measured spectrum has been found.

\section{Initial Doping Experiments}
As mentioned in the introduction, we have made some preliminary
attempts to dope our new tetragonal form of CuO, which like the
other transition metal monoxides, we expect to be a Mott insulator
of the charge-transfer type.\cite{mott1949,zaanen1985} To attempt
charge transfer doping, without breaking vacuum, alkali metals
were deposited on top of our tetragonal CuO films. Alkali metals
were chosen because their outer electrons are very weakly bound
and would be expected to be transferred into the CuO. This
property also makes them very reactive, and so we have not
attempted analyzing these samples outside our deposition chamber.
Of course, even if charge is transferred the question remains
whether the charges will be mobile or localized.  A capability to
measure the transport properties of our samples without leaving
the UHV environment is under development.

\begin{figure}[tb]
\centering
\includegraphics[width=0.5\columnwidth]{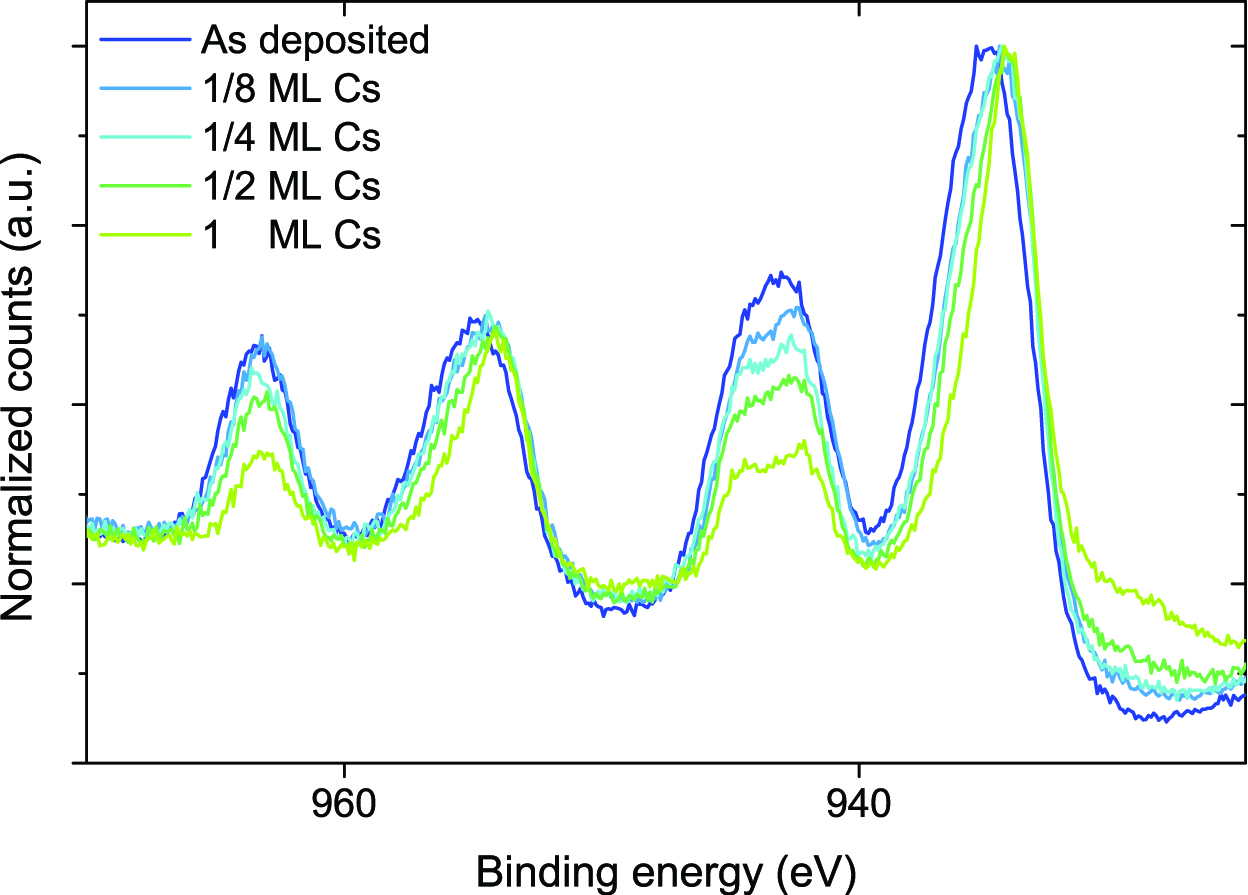}
\caption{XPS Cu 2\emph{p} spectrum of an as-deposited tetragonal CuO film and spectra taken after deposition of
approximately $\frac{1}{8}$, $\frac{1}{4}$, $\frac{1}{2}$, and 1.0 MLs of Cs. The spectrum changes systematically
towards a Cu$^{1+}$ spectrum, which suggests charge carriers are created. The increase of the background at low binding
energy is caused by a nearby Cs peak.} \label{figure7}
\end{figure}

Two alkali metals were used: Li and Cs. The XPS results of one of
the charge transfer doping experiments using Cs are shown in Fig
\ref{figure7}. Deposition was incremental, and after each
deposition photoemission spectra were taken. The Cu 2\emph{p}
spectra in \ref{figure7} show a systematic change with increasing
Cs coverage. The O 1\emph{s} peak at about 530 eV does not change
significantly when Cs is deposited. The Cs peaks themselves are
lower than one would expect for elemental Cs by about 1 eV. CsOH
is 1.4 eV lower than Cs metal, and the lower binding energy
therefore suggests that Cs has donated an electron. The spectral
shift in the 2\emph{p} peaks can be explained by doping of the Cs
electrons into the CuO: in a localized picture this would generate
more Cu$^{1+}$ at the expense of Cu$^{2+}$ and make the spectrum
look more like Cu$_2$O. On the other hand, a similar effect would
be expected to occur if Cs removed oxygen from the CuO, which also
creates more Cu$^{1+}$. In both cases, the CuO would be electron
doped, but UPS measurements showed no Fermi edge, and some
charging was observed during XPS measurements.  Note that we used
the Ti 2$p$ peak from the substrate as an internal energy
reference. The Li doped samples showed a similar trend.

\section{Discussion}

The RHEED and XPD results seem unequivocal that a tetragonal structure of CuO has been synthesized, through the use of
epitaxy. The more interesting question is what are the electronic properties of this higher symmetry phase.  Comparison
of our XPS results for the tetragonal form of CuO with those for tenorite reveals that the main peak becomes broader
and the satellite peak sharper when the tetragonal phase is formed. The intensity distribution in the satellite peak
shifts more to lower binding energies. According to Van Veenendaal and Sawatsky,\cite{veenendaal1993,veenendaal1993b}
the width of the main peak is related to whether the screening electrons come from the closest ligand atoms or from
ligand atoms further away. This would suggest that the screening electrons are more delocalized in the tetragonal
structure. Note that while the shortest Cu--O bond lengths in the tetragonal structure are similar to those in
tenorite, there are also much longer bond lengths in the tetragonal structure, which might have an influence on the
nature of the electrons.  For the construction of the unit cell we have assumed all 180$^{\circ}$ Cu--O--Cu bonds,
whereas they could be less due to Jahn-Teller distortions for example. Other measurements are required to
accurately determine the bond angles, which are known to have a profound impact on especially the magnetic properties.

The structure of the satellite peak is harder to understand. The similarity with CuCl$_2$ spectra from the literature
is striking though.\cite{vanderlaan1981} Okada \emph{et al}.\cite{okada1989} have performed extensive modeling of the
satellite peak of CuCl$_2$. They argue that the shape of the satellite peak is mostly determined by the
coupling-strength ratio between the $\sigma$ and $\pi$ bondings and the amount of hybridization of these states.
Following their model suggests that the degree of hybridization is weaker in the tetragonal structure than in tenorite.
In other words, the bonding in the tetragonal structure seems to have a more ionic character.

The UPS spectrum of tetragonal CuO is very similar to that measured for tenorite. Some notable differences include the
intensity of the low energy peak, the splitting of the main peak, and a general shift of the spectrum to higher binding
energies of about half an eV\@. Delocalized band calculations (local density approximation) performed for tenorite do
not reproduce measurements on the tetragonal phase, suggesting the valence bands are changed with the structural
change. Electron correlation effects are important in these materials, and neglecting them results in incomplete DOS
spectra and conducting ground states. The best predictions come from cluster calculations (configuration interaction
(CI) calculations), which predict the entire spectrum accurately, but shifted to slightly lower binding energy. Eskes
\emph{et al}.\cite{eskes1990} have investigated the nature of the first ionization state of CuO, which has either a
triplet or a singlet character. The singlet state peak is the one closest to the Fermi level (1.9 eV in our spectrum)
and the triplet state is the next one up (3.6 eV for this work). The energy difference between these peaks is closely
related to the Cu--O distances, specifically the ratio between the Cu--O distance out-of-plane and in-plane, which is
about 1.4 for the tetragonal unit cell. Based on their model we would expect to see an energy difference of 0.85 eV
between the two peaks. The measured distance is much larger for reasons that are not well understood. An important next
step in this research will be to specifically calculate the band structure for the tetragonal CuO.

\section{Conclusion}
In conclusion, through the use of epitaxy, a nearly rock salt form of CuO has been synthesized that is now available
for physical study.  Preliminary electron structure studies suggest that the charge transfer gap $\Delta$ is enhanced
in the tetragonal phase and the overall bonding seems more ionic. As a new end member of the 3d transition monoxides,
its magnetic properties should be that of a high $T_N$ antiferromagnet.


This work supported by the DoE with additional support from EPRI. We would like to thank P.M. Grant for many useful
conversations and his continuing interest in this work.  We would also like to acknowledge W.A. Harrison and G.A.
Sawatzky for useful conversations and A. Vailionis for critical help in understanding the RHEED patterns.

\bibliographystyle{apsrev}
\bibliography{biblio}

\end{document}